\documentclass[%
 reprint,
 amsmath,amssymb,
 aps,
pra,
]{revtex4-1}

\usepackage{graphicx}% Include figure files
\usepackage{dcolumn}% Align table columns on decimal point
\usepackage{bm}% bold math

\usepackage{amsfonts,latexsym,fancyhdr,graphicx,bbold}

\DeclareMathOperator{\Tr}{Tr}

\newcommand{\ket}[1]{\lvert #1 \rangle}								%comando ``ket''
\newcommand{\bra}[1]{\langle #1 \rvert}               %comendo bra
\newcommand\bigzero{\makebox(0,0){\text{\huge0}}} %big zero in matrices

\begin{document}

\preprint{APS/123-QED}
\title{Thorough evaluation of GHZ generation protocols using conference key agreement}
\author{Valentina Caprara Vivoli}
\email{capraravalentina@gmail.com}
 \affiliation{QuTech, Delft University of Technology, Lorentzweg 1,
   2628 CJ Delft, The Netherlands}

\begin{abstract}
The generation of GHZ states in quantum networks is a key element for
the realization of several quantum information tasks. Given the
complexity of the implementation of such generation, it is not easy to
find an unambigous proof for an optimal protocol.
Motivated by recent improvements in NV center manipulation, we present
and compare an extensive list of protocols for generating GHZ
states using realistic parameters. Furthermore, in order to establish the goodness of the various protocols, we test them
on a specific application, i.e. conference key agreement. We show
that for an high number of nodes the best protocol is one presented
here for the first time.
\end{abstract}

\maketitle

\section{Introduction}
The generation and storage of GHZ states \cite{GHZ89} in a distributed fashion
would allow the realization of several quantum tasks in quantum
networks, namely reducing communication complexity \cite{Buhrman01,
  Buhrman10}, distributed quantum computation \cite{ Cleve97,
  Grover97, Li15, Li16}, quantum repeaters of second and third
generation \cite{ Muralidharan16, Gottesman99, Muralidharan14,
  Munro12}, and atomic clock synchronization \cite{Komar14}. But it is in quantum cryptography that GHZ states find
their most important applications. Examples of that are quantum secret sharing
\cite{Hillery99}, anonimous state transfer \cite{Christandl05}, and
conference key agreement (CKA) \cite{Epping16}.
From the experimental point of view, impressive improvements have
been done in generating bipartite entanglement in a distributed
fashion with NV centers \cite{Bernier13,Gao15}, and trapped ions
\cite{Hucul15,Delteil16}. It is now possible to generate bipartite
entangled states reaching very high fidelities enabling to successfully test
nonlocality \cite{Hensen15}.
However, little effort has been done so far for the realization of
multipartite entanglement. This is because the high fidelities reached
for bipartite entanglement have been realized at the cost of very low generation
rates. Unfortunately, working with several parties could only worsen this result.
In a previous work \cite{Caprara17}, we have investigated how to generate GHZ states in
a quantum network through one single measurement on
ancillary qubits. There, we have shown that there is an intrinsic bound on
the achievable success probability when one wants to generate
entanglement in a distributed fashion in one single round. In the case of bipartite
entanglement, this bottleneck can be overcome through the use of
distillation procedures like the extreme-photon-loss (EPL) protocol \cite{Li15}, that has recently been
experimentally realized \cite{Kalb17}. Unfortunately, this happens at the cost
of decreasing the fidelity. Hence, it is of primary
importance to investigate different protocols in order to get the best
compromise between fidelity and generation rate. This task is not easely doable
since it is not clear how to evaluate the goodness of such a
compromise.
\begin{figure}[h]
\begin{center}
  \includegraphics[width=250pt]{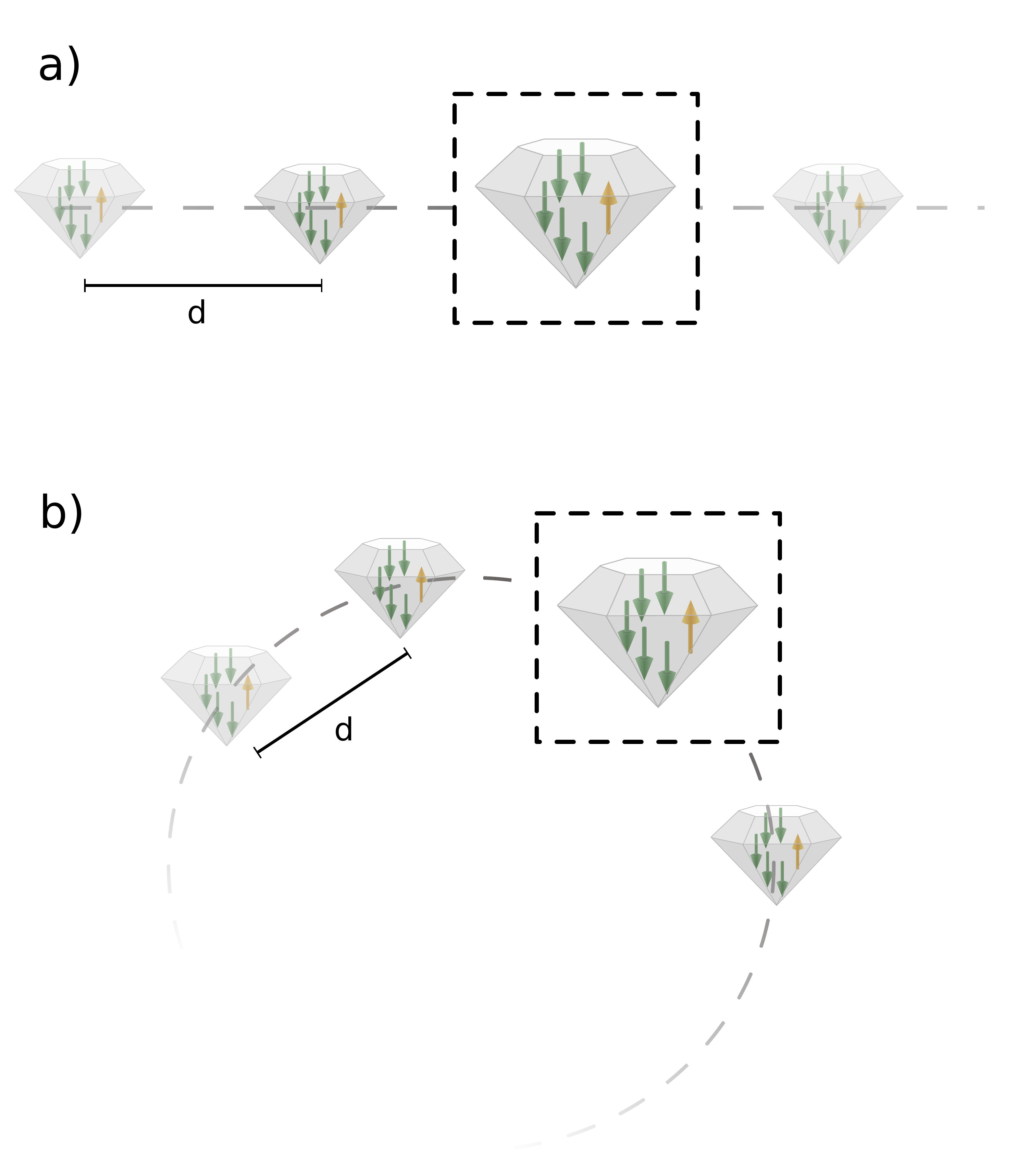}
\caption{Quantum network architectures. Each node is constituted by a
  NV center where one electronic spin (yellow arrow) and up to five nuclear
  spins (green arrows) can be stored. In the case of a) (b)) a linear
  (circular) architecture is represented. In both cases the nodes are distributed at a fixed distance $d$. \label{Architectures}}
\end{center}
\end{figure}
 An approach to the issue is to compare the different protocols in terms of a
specific application and to evaluate the total application rate. Since
the application rate depends both on the generation rate and the
goodness of the multipartite state, it constitutes an unambigous
parameter for selecting a successful protocol. A recent
work \cite{Ribeiro17} analyzes conference key agreement (CKA) in
presence of losses and gives an expression for the asymptotic rate in
a fully device-independent scenario. In this paper, we investigate how to generate GHZ states between nearby
nodes through distillation procedures, error correction, and linear
optics.
\begin{figure}[h]
\begin{center}
  \includegraphics[width=150pt]{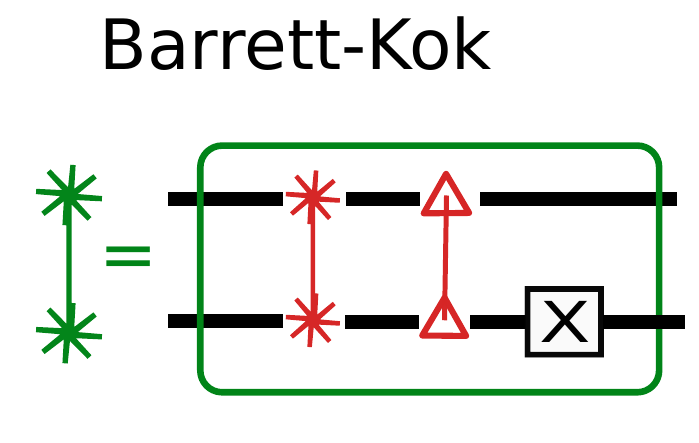}
\caption{Barrett-Kok circuit. An EPR pair is generated between two
  distant nodes through two successful consecutive measurements on
  ancillary modes. Finally, an X rotation is applied on one qubit,
  passing from $\ket{\Psi^+}$ to $\ket{\Phi^+}$. The two red star
  symbol (two red triangle symbol) represents a successful Bell
  measurement over ancillary modes entangled with the $\ket{0}$-levels ($\ket{1}$-levels).}
\end{center}
\end{figure}
\begin{figure}
\begin{center}
  \includegraphics[width=180pt]{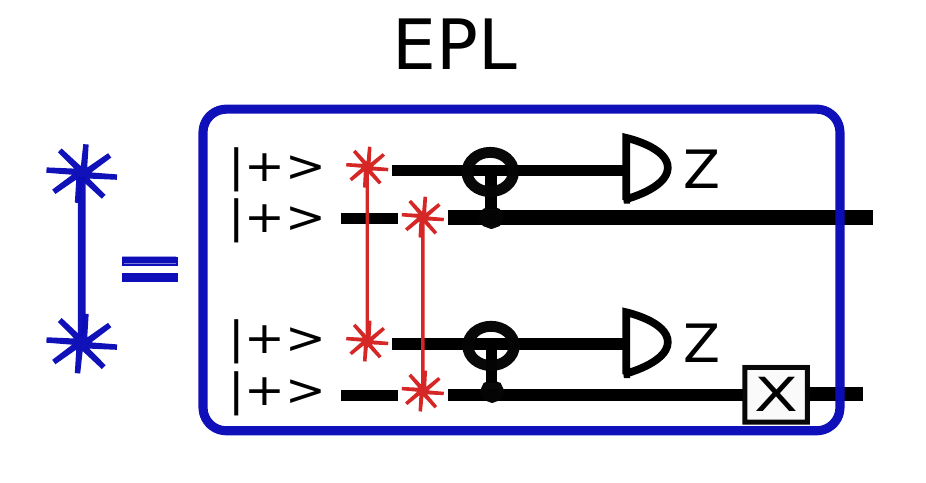}
\caption{EPL circuit. An EPR pair is generated between distant nodes
  through a distillation procedure. Firstly, two non-maximally
  entangled pairs are generated through two distinct successful Bell
  measurements. Secondly, CNOT and measurements are performed
  locally. Finally, an X rotation is applied on one qubit.}
\end{center}
\end{figure}
\begin{figure}[h]
  \begin{center}
  \includegraphics[width=230pt]{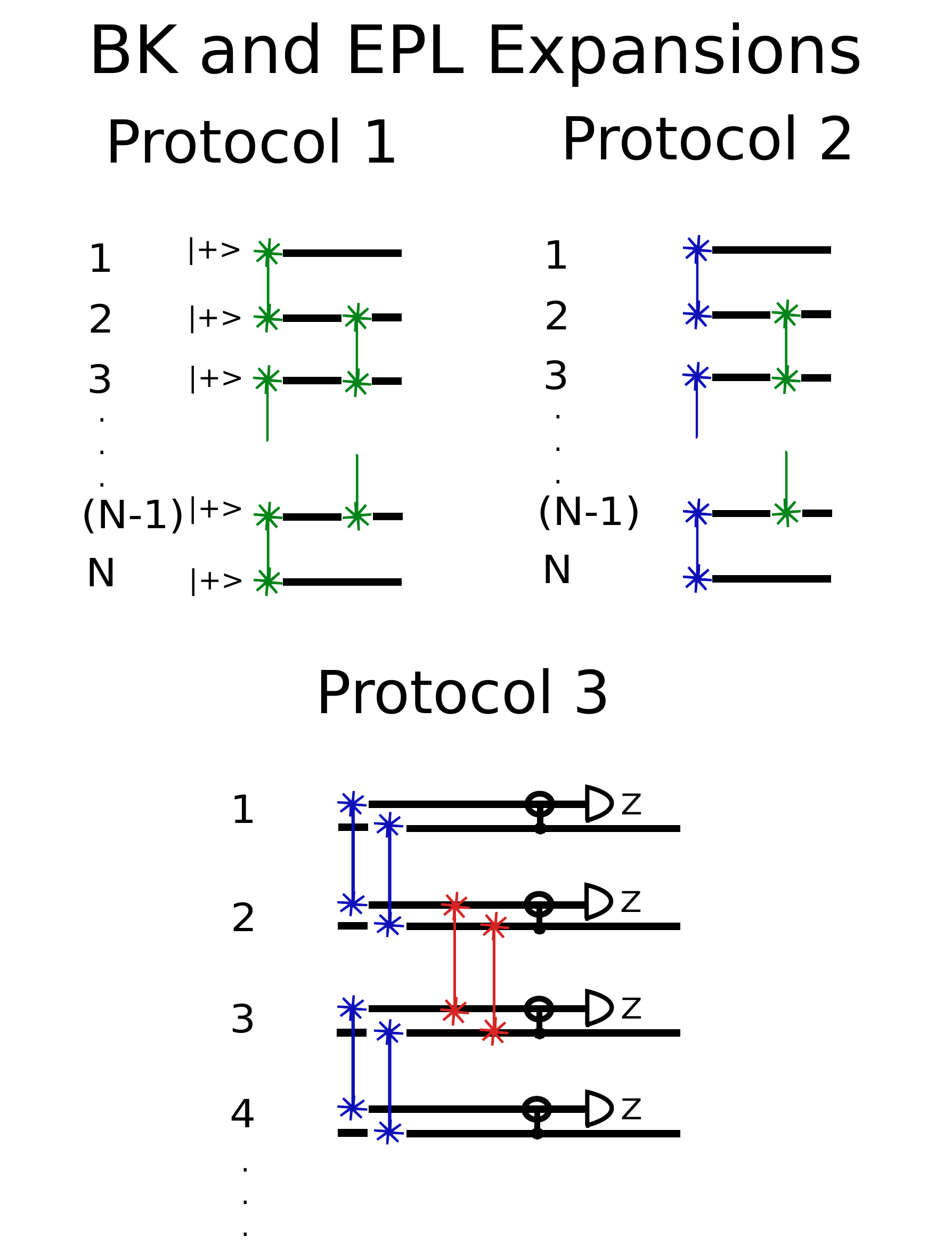}
\caption{Circuit diagrams for the linear protocols that are BK/EPL
  expansions. Protocol 1: the BK protocol (two green star symbols) is
  applied in two subsequently time steps
  between neighbouring nodes. Protocol 2: $N/2-1$ successful Barrett-Kok
    procedures are performed between $N/2$ EPL pairs. Protocol 3: a distillation procedure
  (CNOT and local measurement) is
  applied on two non-maximally multipartite entangled states. \label{LinearProt1}}
  \end{center}
    \end{figure}
\begin{figure}[h]
  \begin{center}
\includegraphics[width=230pt]{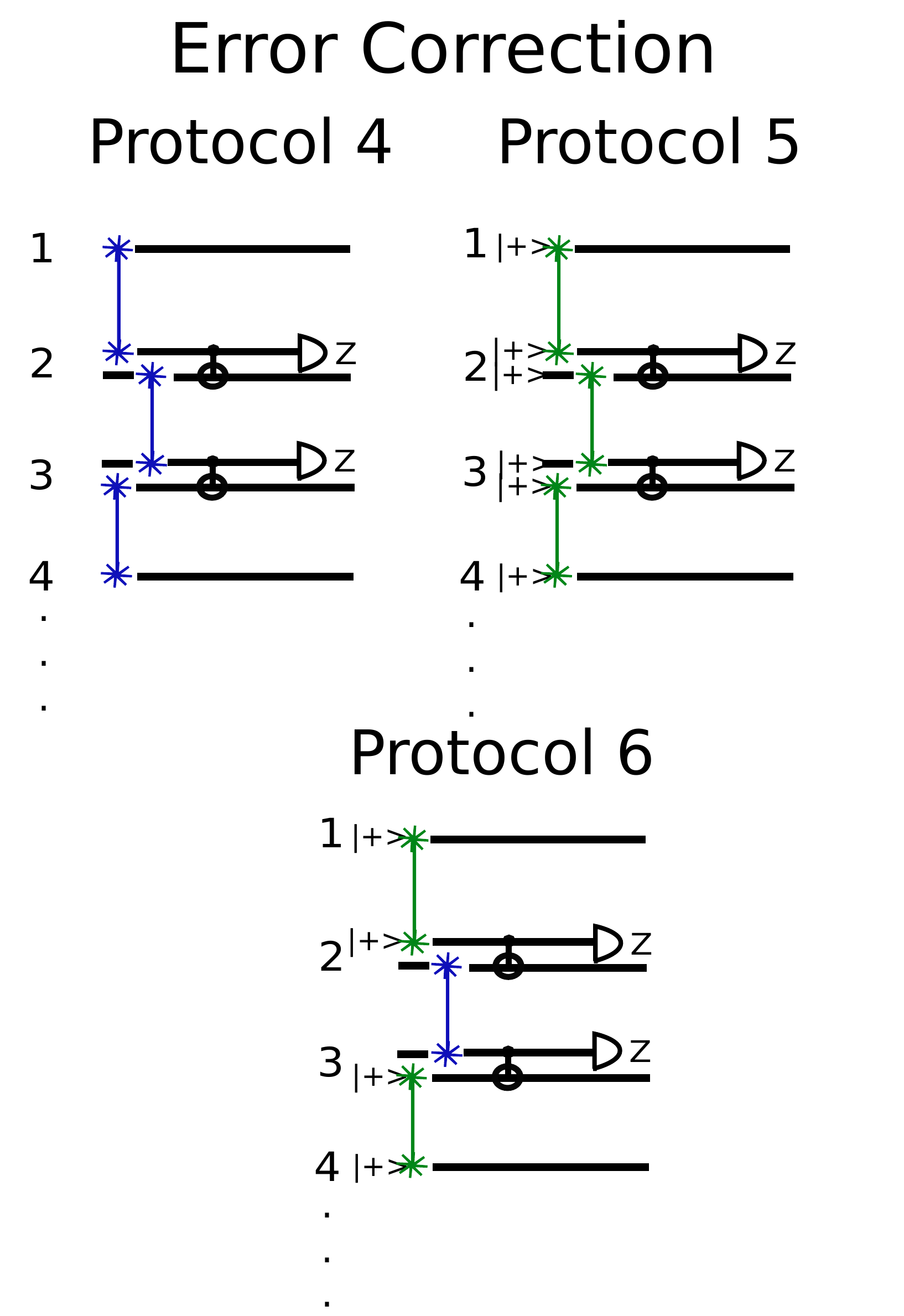}
\caption{Circuit diagrams for the linear protocols that exploit
  bipartite entanglement and error correction. Protocol
  4: CNOTs and measurements are locally applied over $(N-1)$ EPL
  pairs. Protocol 5: CNOTs and measurements are locally applied over
  $(N-1)$ Barrett-Kok pairs.  Protocol 6:
    an error correction procedure is performed over $N/2$ Barrett-Kok and
    $(N/2-1)$ EPL pairs.\label{LinearProt2}}
  \end{center}
   \end{figure}
\begin{figure}[h]
  \begin{center}
 \includegraphics[width=230pt]{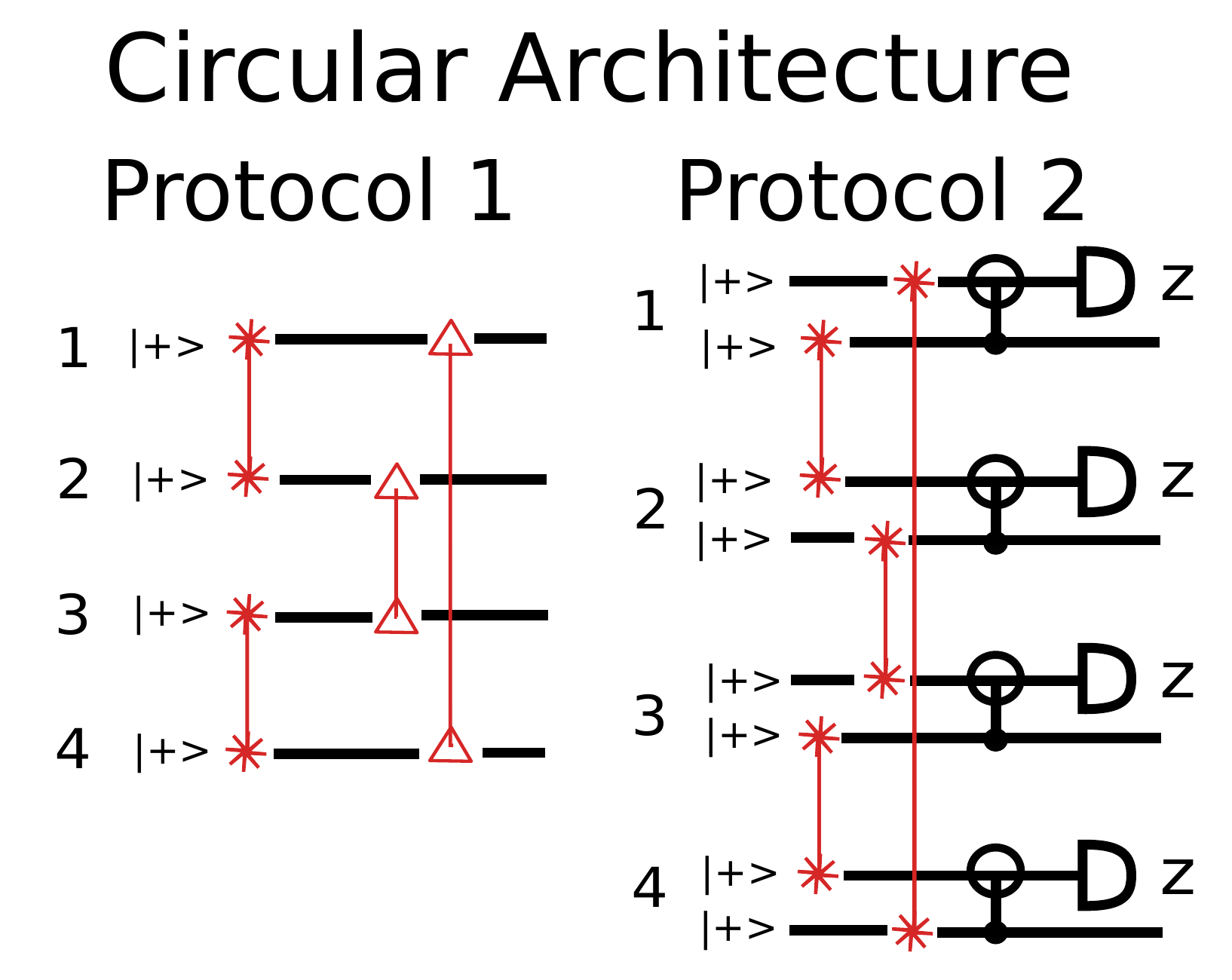}
  \end{center}
  \caption{Circuit diagrams for circular protocols 1 and 2. Protocol
    1: $N$ successful measurements are performed over $N$ nodes in a circle. Protocol 2: a distillation procedure consisting on a
    CNOT and a measurement in each node, is performed over
    non-maximally entangled pairs.\label{CircProt}}
  \end{figure}
The system we envision is composed by N nodes, each one containing
up to five qubits, even though at most two are necessary. One and two
qubit logical operations can be performed locally. The nodes interact between each other through
ancillary photonic modes that are entangled with the qubit
levels. Since the implementable protocols depend on the configuration
the nodes are arranged in,
 we consider two different architectures, or along a
line, or along a circle. Notice that in the latter configuration each
node is close enough to only two nodes, such that it is easy for it to
generate an entangled pair with each one of them. Here, we first
present and compare eight protocols in terms of fidelity and
generation rate in a realistic scenario, namely NV centers.
For linear architectures, we present six protocols,
regrouped in two subsets. One set (Fig. \ref{LinearProt1}) is composed by protocols that
consist in applying repeatedly the Barrett-Kok (BK)\cite{Barrett05, Lim05} and
EPL \cite{Li15} bipartite techniques. The other set
(Fig. \ref{LinearProt2}) is composed by all the protocols where, firstly,
maximally entangled pairs are realized between nearby nodes, and,
secondly, error correction is used to generate the final
multipartite entangled state. This approach has
already been proposed by Komar, et al. \cite{Komar16}. We, here, further
investigate this possibility varying the way the maximally
entangled pairs are generated. For a circular architecture, we present
two protocols (Fig. \ref{CircProt}), one already envisioned in
\cite{Caprara17}, and a new distillation one.
As a term of comparison, we use the
minimal required fidelity for asymptotic CKA
\cite{Ribeiro17}. We, furthermore, derive the total asymptotic rate for CKA for
all the reasonable protocols. The results show that there is a clear
trend as the number of nodes increases. Indeed, for high number of
nodes, the circular protocols reveal to be dozens of orders of
magnitude faster than the linear protocols. The cause of that has to
be sought in the possibility of connecting each node with other two
nodes. As a consequence the number of probabilistic operations
necessary to generate maximally entangled states is highly reduced.
One ends up with GHZ states low decohered and high generation rates.

\section{Modeling NV centers and losses}
In order to evaluate the different protocols in presence of loss and decoherence we need to contextualize
them choosing a specific system. NV centers are the perfect candidates
for such protocols. In this section, we describe the error model
for NV centers. In the system that we envision, each node is
constituted by an NV center. For the sake of simplicity, we assume the
number of nodes $N$ to always be even. For each NV center we have at
our disposal several spins, namely an electronic spin and up to five nuclear
spins \cite{Reiserer16}. Only the
electronic spin can directly be manipulated and any operation on
nuclear spins is performed through the electronic spin. We assume that
any operation on a single nuclear spin is not affected by decoherence.
When one access the electronic spin, the nuclear spins undergo
dephasing due to hyperfine interaction between the first and the
second ones \cite{Reiserer16}. The expression for a dephasing channel on a
density matrix $\rho$ is
the following
\begin{equation}
D_{\text{deph}}(\rho)=\frac{1+\lambda}{2} \rho+\frac{1-\lambda}{2} \sigma_z \rho \sigma_z,
\end{equation}
where $\lambda=e^{-a n}$ quantifies the noise. In the
expression of $\lambda$, $n$ is the number of
attempts that have been performed on the electronic spin, while $a$ depends both on the attempt
of accessing the other qubit in the same node and the time required
for performing the specific operations. The expression for $a$ is
$a=a_0+a_1   t_{\text{step}}$\cite{Rozpedek17},
where $a_0=\frac{1}{2000}$ per attempt, $a_1=\frac{1}{3}$ per second
due to the storing time, and $t_{\text{step}}$ is the time required to
perform the specific step of the protocol. When the decohering nuclear
spin is not stored in a NV center, where one is operating on the
electronic spin, $a$ takes the form $a=a_1   t_{\text{step}}$. Each $\lambda$ factor must be averaged over the number of attempts, i.e.
\begin{equation}\begin{split}
\langle\lambda\rangle&=\frac{\sum_{n=0}^{\infty}\left[P_{\text{succ}}
    (1-P_{\text{succ}})^n e^{-a
      n}\right]}{\sum_{n=0}^{\infty}\left[P_{\text{succ}}
    (1-P_{\text{succ}})^n \right]}\\
&=\frac{P_{\text{succ}} e^{a}}{e^{a}-1+P_{\text{succ}}},
\end{split}\end{equation}
where $P_{\text{succ}}$ is the probability of success per attempt of the specific
operation, and the sums of the series are performed over
all the attemps.
In all the protocols the terms outside the space $\{\ket{0}^{\otimes
  N},\ket{1}^{\otimes N}\}$, where N is the number of nodes, are
nullified. As a consequence, the final density matrix $\rho_{\text{final}}$ takes the form
\begin{equation}\rho_{\text{final}}=\frac{1}{2}\left(
\begin{array}{ccccc}
1&0&\cdots&0&\langle\lambda\rangle_{\text{tot}}\\
0&&&&0\\
\vdots&&\bigzero&&\vdots\\
0&&&&0\\
\langle\lambda\rangle_{\text{tot}}&0&\cdots&0&1
\end{array}\right),\label{GHZdeph}
\end{equation}
where $\langle\lambda\rangle_{\text{tot}}=\prod_i \langle\lambda_i\rangle$ is the product between all
the factors $\langle\lambda_i\rangle$ that cause decoherence during
the protocol on all the spins.
The losses in the optical setup are represented through the total
transmittivity $\eta$, i.e.
\begin{equation}
\eta=\eta_D p_{\text{fc}}p_{\text{out}}10^{-\frac{\alpha d}{L_0}},
\end{equation}
where $\eta_D=1$ is the detector efficiency, $p_{\text{fc}}=0.3$ is the
frequency conversion efficiency, $p_{\text{out}}=0.3$ is the NV
outcoupling efficiency, $L_0=20$ km is the attenuation length of
the fibres \cite{vanDam17}, and $d$ is the distance between two
neighbouring nodes.

\section{Fidelity and generation rate}
\begin{figure}[h]
  \begin{center}
  \includegraphics[width=225pt]{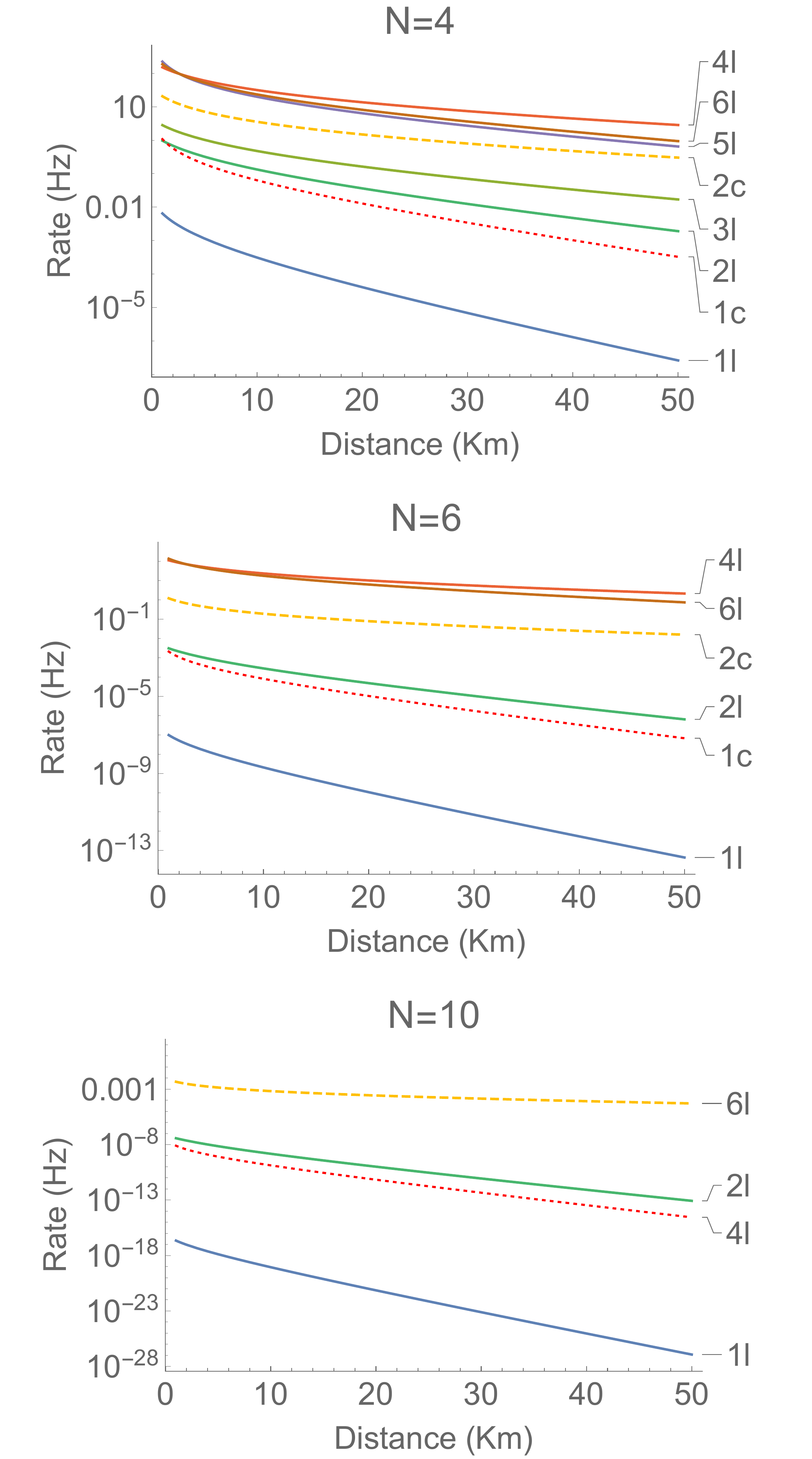}
  \end{center}
  \caption{Generation rate for $N=4$, $6$, and $10$ for the different
    protocols. The rate of generation $R_{\text{GHZ}}$ as a function
    of the distance is plotted for all the protocols that have
    fidelity above the thresholds.\label{RateTot}}
  \end{figure}

In order to estimate the goodness of each protocol it is useful to
evaluate the fidelity of the final state for each protocol with the
GHZ state. We, furthermore, compare the fidelities with the minimal
required fidelity for CKA with the protocol presented in
\cite{Ribeiro17}. In the aforementioned work, the system is affected by
depolarizing noise, i.e.
\begin{equation}
D_{\text{depol}}(\rho^{\text{qubit}})=(1-p)\rho^{\text{qubit}}+p \Tr(\rho^{\text{qubit}})\frac{\mathbb{1}}{2},
\end{equation}
where $p$ is the noise that affects each spin.
The expression for the fidelity with the GHZ state (see appendix \ref{FidCKA}) is
\begin{equation}
F_{\text{CKA}}=\frac{1}{2}\left(1-\frac{p}{2}\right)^N+\frac{(1-p)^N}{2}+\frac{1}{2}\left(\frac{p}{2}\right)^N,\label{FCKA}
\end{equation}
where $N$ is the number of nodes. The maximal $p$ for each $N$
in order to achieve a
positive CKA rate is numerically evaluated in \cite{Ribeiro17}.
The fidelity between a GHZ state and a state in
the form of Equ. \eqref{GHZdeph} is
\begin{equation}
F=\frac{1+\langle\lambda\rangle_{\text{tot}}}{2}.
\end{equation}
the details of both $\langle\lambda\rangle_{\text{tot}}$ and generation rate for all the protocols
are given in appendix \ref{DephFact} and \ref{GenRate}.
\begin{figure}[h]
  \begin{center}
  \includegraphics[width=250pt]{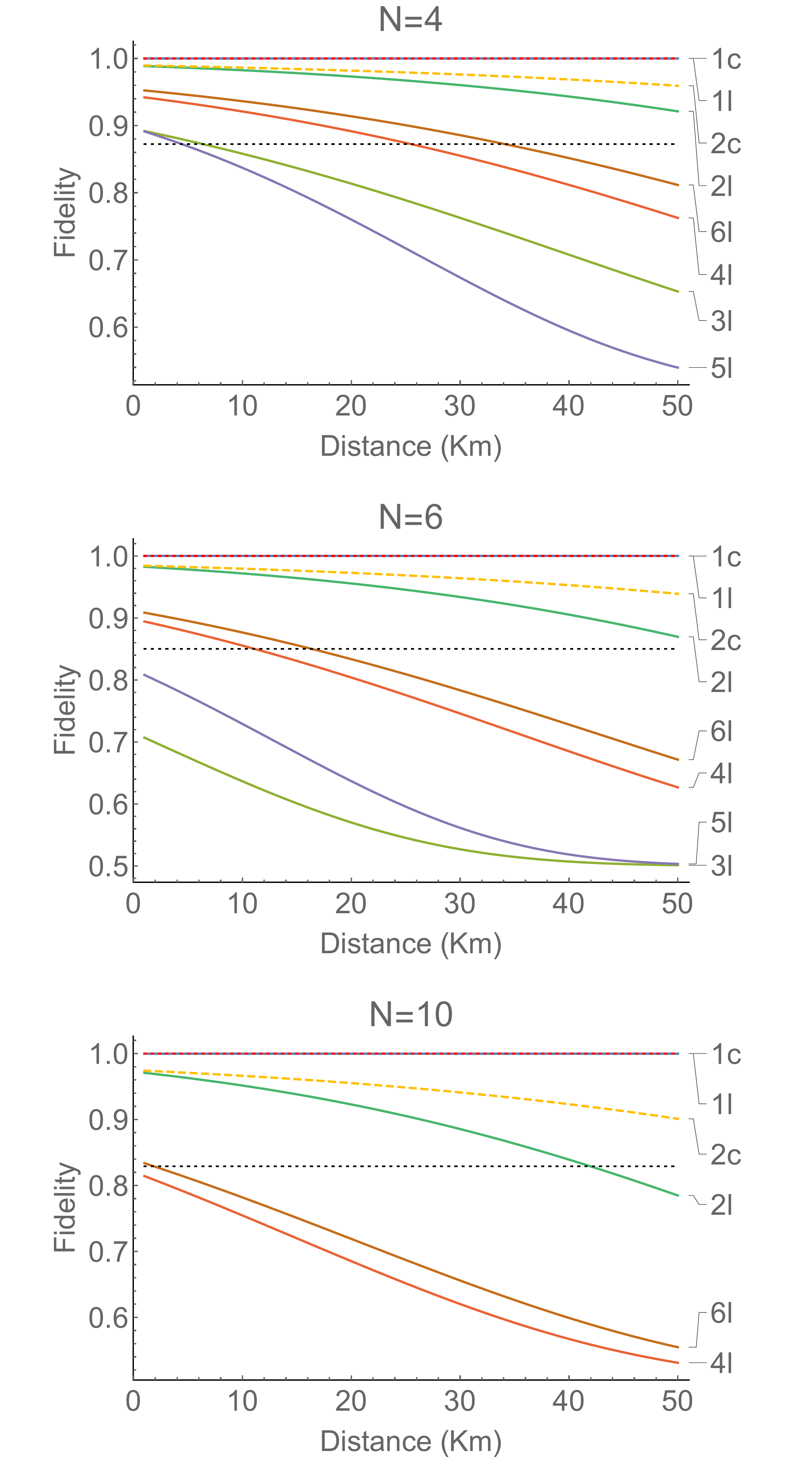}
  \end{center}
  \caption{Fidelity for $N=4$, $6$, and $10$ for the different protocols. The
    fidelity of the final state with the GHZ as a function of the
    distance is plotted for all the protocols. The black line
    represents the minimal fidelity necessary to implement the CKA
    device-independently. For $N=10$, linear protocol 3 and 5 have
    been omitted.\label{FidTot}}
  \end{figure}
The fidelities for $N=4$, $N=6$ and $N=10$ as a function of the distance
between the nodes are plotted in Figs. \ref{FidTot}. The minimal fidelity for CKA is the
dashed black line. For $N=4$, all the protocols are above the
threshold for some range. However, linear protocols do perform
worse. Specifically, it seems that error correcting protocols present
more decoherence. Increasing the number of nodes, linear protocols
3,4,5, and 6 become useless. Hence, only linear and circular protocols
1s and 2s are resistant to decoherence.
The rates for only the successful protocols as a function of the distance are plotted in
Figs. \ref{RateTot} for $N=4$, $N=6$ and $N=10$. From the plots,
it is unclear what protocol is the most advantageous among the
several.
The linear protocols 4, 5, and 6 are not usable when the
distance between nodes is above around 30 km. However, the latest
consideration does not exclude that they are the most suitable for a
short distance. Indeed, the linear protocols 4, 5, and 6 have better rates than
the others, but have high decoherence.
In the next section, we are going to discuss how to overcome this
difficulty using the CKA asymptotic key rate.

\section{Conference key agreement rate\label{CKA}}
\begin{figure}[h]
  \begin{center}
 \includegraphics[width=225pt]{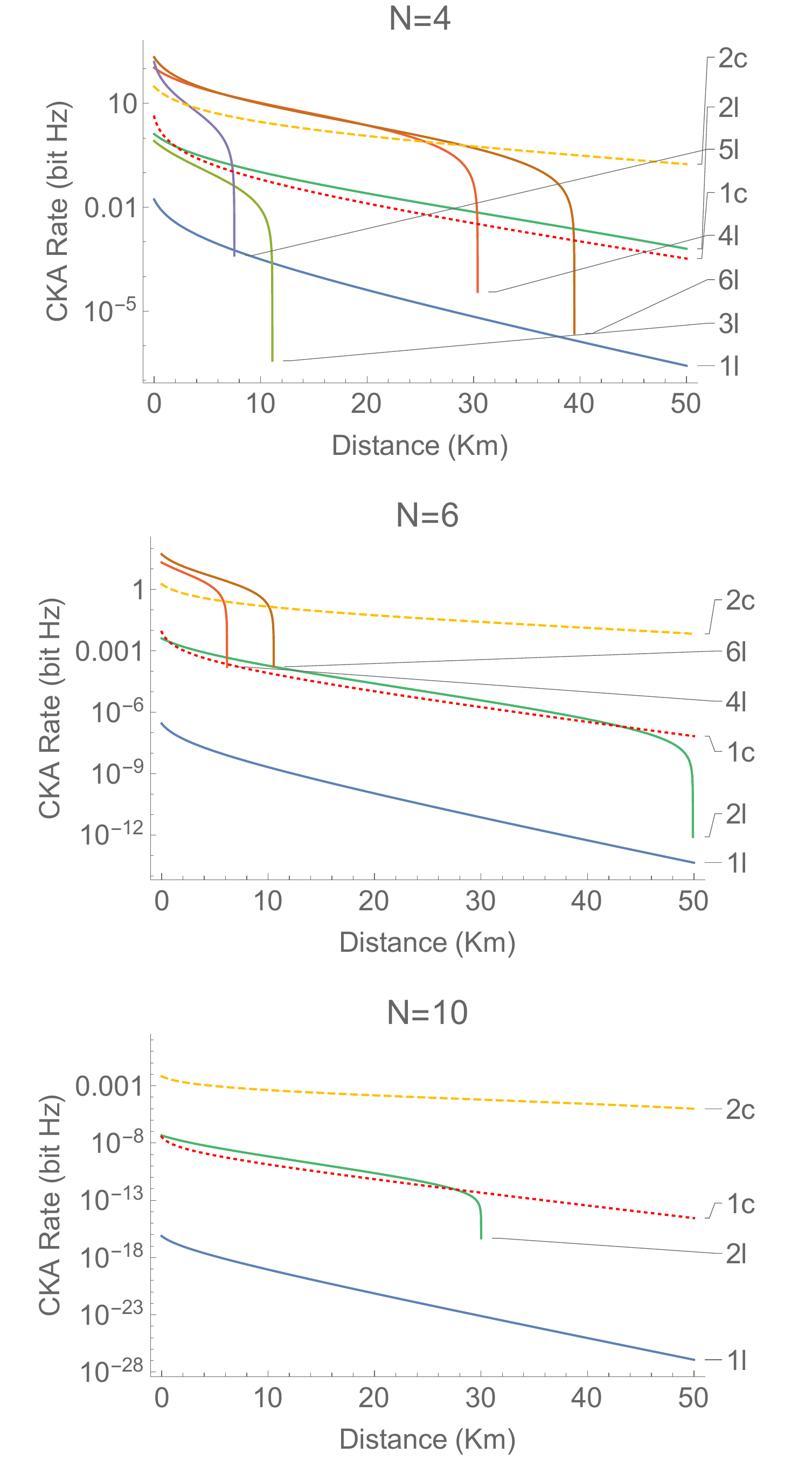}
  \end{center}
  \caption{Application rate for $N=4$, 6, and 10 for different
    protocols. The CKA rate is plotted as a function of the distance
    between two nodes for all the protocols that have a sufficient
    high fidelity. In several cases, CKA rates
    suddenly drops to 0 when the dephased GHZ state does not anymore
    violate the MABK inequality.\label{CKARateTot}}
  \end{figure}

In this section, we calculate the CKA asymptotic key rate starting
from the expression of the fidelity and generation rate for a given protocol, and analyze
it as a function of the distance between the nodes.
The expression for the CKA asymptotic key rate is \cite{Ribeiro17}
\begin{equation}
\tilde{R}_{\text{CKA}}=1-h\left(\frac{1}{2}+\frac{1}{2}\sqrt{\frac{MK_N^2}{2^{N-2}}-1}\right)-h(Q),
\end{equation}
where $h(\cdot)$ is the binary entropy, $MK_N$ is the MABK value
\cite{Mermin90,Ardehali92,Belinski93} in the N-partite case for
the dephased GHZ state, and $Q$ is the quantum bit error rate
(QBER). The QBER is given by the probability of getting a flip
error. Hence, in the case of $\rho_{\text{final}}$ the QBER is 0,
i.e. $Q=0$. We have then,
\begin{equation}
h(Q=0)=\lim_{Q\rightarrow 0}\left[Q\log_2Q-(1-Q)\log_2(1-Q)\right]=0.
\end{equation}
It can be numerically proven that the $MK_N$ violation for
$\rho_{\text{final}}$ (Equ. \eqref{GHZdeph}) is
$2^{\frac{N-1}{2}}\langle\lambda\rangle_{\text{tot}}$. Thus, the CKA
asymptotic key rate becomes
\begin{equation}
\tilde{R}_{\text{CKA}}=1-h\left(\frac{1}{2}+\frac{1}{2}\sqrt{2 \langle\lambda\rangle_{\text{tot}}^2-1}\right).\label{RateCKA}
\end{equation}
The rate in Equ. \eqref{RateCKA} must be multiplied by the GHZ generation rate
$R_{\text{GHZ}}$, i.e.
\begin{equation}
R_{\text{CKA}}=\tilde{R}_{\text{CKA}}R_{\text{GHZ}}.
\end{equation}
The results for $N=4$, $6$ and $10$ as a function of the distance are shown in
Figs. \ref{CKARateTot}.
For 4 and 6 nodes, the error correcting protocols are the most
effective for short ranges ($\approx 40$ km and $20$, respectively). For longer
ranges, the circular protocol 2 is the fastest one. Concerning the
linear protocols, albeit extremely slow, protocols 1 and 2 still perform.

\section{Conclusion}
In this article, we have reported of a detailed study of several
protocols for GHZ generation in a quantum network composed by NV
centers. We evaluate the effectiveness of these
protocols through the calculation of three values, the fidelity, the
generation rate, and a figure of merit that combines both
fidelity and generation rate, i.e. the asymptotic CKA rate. Indeed,
the fidelity and the generation rate are common and widespread
measures of the goodness of any protocol and are easy to read and
interpret for a great audience. However, we found that such a complex
protocol analysis was incomplete, since the two observables vary
independently from each other. Testing protocols over a specific
application is not new \cite{Nickerson14}. What it is new is the
extensiveness of the study, both in the variety of the protocols and
the decoherence analysis. Concerning the protocols, some of them have been proposed in recent papers
\cite{Caprara17,Komar16}, some have partially been readapted from previous work
\cite{Barrett05, Lim05}; finally, only circular protocol 2 is
completely new. In any case, they entirely cover the approaches so-far
envisioned.
Concerning the decoherence analysis, our study differs from all the
previous ones, since we have considered a realistic scenario,
including the decoherence due to waiting time, that reveils to be
critical for the effectiveness of the protocols.
The results show
that as the number of nodes increases, the best protocol is the one
proposed for the first time, here. Our interpretation is that, for the
new protocol, the time required for the generation of the
intermediate entanglement is extremely low, resulting in few
decoherence and relatively high generation rate. However, this is
possible only for circular architectures and not linear. Therefore,
there are doubtless cases when such a protocol can not be
implemented because of the network architecture. In this instance, the
best protocol is the linear protocol 1, i.e. a protocol consisting of
only one round. Moreover when one focus only on linear architectures,
surprisingly, only protocols 1 and 2, protocols with very low rates,
are available. On the contrary, all the protocols that extensively use
distillation and error correction result to be too noisy for CKA. This
counterintuitive result is a direct consequence of the decoherence due
to the waiting times between bipartite entanglement generation and the
following step.
Concerning the decoherence analysis, few remarks have
to be done. First, it is important to stress that the system might encounter
other decoherence processes, for example depolarizing
channels. Nevertheless, we notice that we have compared the
fidelities with a trademark fidelity computed for depolarizing
noise. In that frame, some protocols showed to not have enough
good fidelities for some specific distances. We have found the same
result in section \ref{CKA} for the same distances analyzing the CKA rate. It is, then, our
understanding that the qualitative results do not significantly change
depending on the decoherence nature. Secondly, we do aknowledge that
few sources of decoherence and imperfections have not been taken into
account. Examples are the detector dark counts, and the unfidelities
of one, two qubit logical ports. Nevertheless, they seem to be of
lower impact on the results and, then, do not affect our conclusion on
the analysis.  Further research should be in two directions. the first is
to test how well other quantum
information tasks that exploit multipartite entanglement perform with
these protocols. The second is to focus on finding new alternative
linear protocols that can improve the applications performance.
 We want, then, to conclude saying
that our study represents a detailed and realistic work on GHZ
generation that reveils some misconceptions on the main network noise
sources and proposes a new promising protocol.
\\
\\

\textbf{Aknowledgements}\\
We thank J\'{e}r\'{e}my Ribeiro, Filip Rozp\c{e}dek, and Stephanie Wehner for fruitful
discussions. %We acknowledge support from the Netherlands Organisation
%for Scientific Research (NWO) through a VIDI grant.

\newpage

\textbf{\Large Appendix}

\section{Calculation of the fidelity for depolarizing noise \label{FidCKA}}
\hspace{.5cm}
In \cite{Ribeiro17}, a depolarizing channel $D_{\text{depol}}(\rho^{\text{qubit}})
$ acts on each qubit. The expression of $D_{\text{depol}}(\rho^{\text{qubit}})$ is
\begin{equation}
D_{\text{depol}}(\rho^{\text{qubit}})=(1-p)\rho^{\text{qubit}}+p \Tr(\rho^{\text{qubit}})\frac{\mathbb{1}}{2},
\end{equation}
where $\rho^{\text{qubit}}$ is the density matrix of a single qubit
and $p$ is the depolarizing factor. The total state $\tilde{\rho}_{\text{final}}$ is, then
\begin{equation}
\tilde{\rho}_{\text{final}}=D^{\otimes N}(\ket{\text{GHZ}}\bra{\text{GHZ}}_N),
\end{equation}
where $N$ is the number of nodes. We need to rewrite $\tilde{\rho}_{\text{final}}$ in a more
explicit way, i.e.
\begin{equation}\begin{split}
\tilde{\rho}_{\text{final}}&=(1-p)^N \ket{\text{GHZ}}\bra{\text{GHZ}}_N\\
&+\sum_{S\in \{0,1\}^N,S\not= \vec{0}}\Tr_S(
\ket{\text{GHZ}}\bra{\text{GHZ}}_N)\\
&\otimes\frac{\mathbb{1}}{2^{W_H(S)}}(1-p)^{N-W_H(S)}p^{W_H(S)}\\
&=(1-p)^N \ket{\text{GHZ}}\bra{\text{GHZ}}_N\\
&+\sum_{S\in \{0,1\}^N,
S\not=
\vec{0}}\frac{\ket{0}\bra{0}^{N-W_H(S)}+\ket{1}\bra{1}^{N-W_H(S)}}{2}\\
&\otimes \frac{\mathbb{1}}{2^{W_H(S)}}(1-p)^{N-W_H(S)}p^{W_H(S)},
\end{split}\end{equation}
where $W_H(S)$ is the hamming weight of the vector $S$.
The fidelity with the $\ket{\text{GHZ}}$ as a function of $p$ is
\begin{equation}\begin{split}
F&=\Tr(\ket{\text{GHZ}}\bra{\text{GHZ}}_N\tilde{\rho}_{\text{final}})\\
&=(1-p)^N +\left(\frac{p}{2}\right)^N\\
&+\frac{1}{2}\sum_{S\in \{0,1\}^N,
S\not=
\vec{0},\vec{1}}\frac{1}{2^{W_H(S)}}(1-p)^{N-W_H(S)}p^{W_H(S)}\\
&=(1-p)^N+\left(\frac{p}{2}\right)^N+\frac{1}{2}\sum_{k=1}^{N-1}\binom{N}{
k}(1-p)^{N-k}\left(\frac{p}{2}\right)^{N}\\
&=\frac{1}{2}\left(1-\frac{p}{2}\right)^N+\frac{(1-p)^N}{2}+\frac{1}{2}\left(\frac{p}{2}\right)^N.
\end{split}\end{equation}
$p$ is numerically evaluated in \cite{Ribeiro17}.

\section{Calculation of the dephasing factors and fidelities \label{DephFact}}
In this section, we first derive twelve dephasing terms
$\langle\lambda_i\rangle$ intervening during the protocols in single
nodes. Afterwards, we report the expressions of all the fidelities for
the different protocols. In the main text, we derived the expression
for $\langle\lambda_i\rangle$, i.e.
\begin{equation}
\langle\lambda_i\rangle=\frac{P^i_{\text{succ}} e^{a^i}}{e^{a^i}-1+P^i_{\text{succ}}}.
\end{equation}
Each $\langle\lambda_i\rangle$, that depends on $P^i_{\text{succ}}$
and $a^i$, gives account of the dephasing of a single nuclear spin
involved in the protocols. Concerning the expression of $a^i$, it
depends on the two terms $a_0$ and $a_1$. $a_1$ gives count of the
decoherence of the nuclear spin while it has to be stored, while $a_0$
represents the decoherence caused by each attempt of access on the
electronic spin in the same NV center. As a consequence, $a_0$ is not
present in $a^i$ when the decohering nuclear spin is not in a NV
center whose electronic spin is manipulated.
In Table 1, we report the different processes that occur during the
protocols, the dephasing processes connected to them, the
$P^i_{\text{succ}}$, and the $a^i$.
The $\langle\lambda_i\rangle$ are reported below,
\begin{equation}
\langle\lambda_1\rangle=\frac{P^1_{\text{succ}}  e^{a_0+a_1 t_1}}{e^{a_0+a_1 t_1}+P^1_{\text{succ}} -1},
\end{equation}
\begin{equation}
\langle\lambda_2\rangle=\prod _{l=1}^{\frac{N}{2}-1} \left(\frac{P^2_{\text{succ}} e^{a_1 t_2}}{H_{l} \left(e^{a_1 t_2}+\frac{P^2_{\text{succ}} }{H_{l}}-1\right)}\right)^2,
\end{equation}
\begin{equation}
\langle\lambda_3\rangle=\frac{P^3_{\text{succ}} e^{a_0+a_1
    t_1}}{e^{a_0+a_1 t_1}+2^{2-N}P^3_{\text{succ}}-1},
\end{equation}
\begin{equation}
\langle\lambda_4\rangle=\prod _{l=1}^{\frac{N}{2}} \left(\frac{P^2_{\text{succ}} e^{a_0+a_1  t_2}}{\left(H_{\frac{N}{2}}-H_{l-1}\right) \left(e^{a_0+a_1  t_2}+\frac{P^2_{\text{succ}}}{H_{\frac{N}{2}}-H_{l-1}}-1\right)}\right)^2,
\end{equation}
\begin{equation}
\langle\lambda'_4\rangle=\prod _{l=1}^{\frac{N}{2}-1} \left(\frac{P^2_{\text{succ}} e^{a_1  t_2}}{H_{l} \left(e^{a_1 t_2}+\frac{P^2_{\text{succ}}}{H_{l}}-1\right)}\right)^2,
\end{equation}
\begin{equation}
\langle\lambda_5\rangle=\prod _{l=1}^{\frac{N}{2}-2} \left(\frac{P^2_{\text{succ}} e^{a_1  t_2}}{H_{l} \left(e^{a_1  t_2}+\frac{P^2_{\text{succ}}}{H_{l}}-1\right)}\right)^2,
\end{equation}
\begin{equation}
\langle\lambda_6\rangle=\left(\frac{P^2_{\text{succ}} e^{a_1  t_2}}{H_{\frac{N}{2}-1} \left(e^{a_1  t_2}+\frac{P^2_{\text{succ}}}{H_{\frac{N}{2}-1}}-1\right)}\right)^2,
\end{equation}
\begin{equation}
\langle\lambda_7\rangle=\prod _{l=1}^{\frac{N}{2}-1} \left(\frac{P^2_{\text{succ}} e^{a_0+a_1  t_2}}{\left(H_{\frac{N}{2}-1}-H_{l-1}\right) \left(e^{a_0+a_1  t_2}+\frac{P^2_{\text{succ}}}{H_{\frac{N}{2}-1}-H_{l-1}}-1\right)}\right)^2,
\end{equation}
\begin{equation}
\langle\lambda'_7\rangle=\prod _{l=1}^{\frac{N}{2}-2} \left(\frac{P^2_{\text{succ}} e^{a_1  t_2}}{H_{l} \left(e^{a_1  t_2)}+\frac{P^2_{\text{succ}}}{H_{l}}-1\right)}\right)^2,
\end{equation}
\begin{equation}
\langle\lambda_8\rangle=\prod _{l=1}^{\frac{N}{2}-1} \left(\frac{P^8_{\text{succ}} e^{a_0+a_1 t_1}}{\left(H_{\frac{N}{2}-1}-H_{l-1}\right) \left(e^{a_0+a_1  t_1}+\frac{P^8_{\text{succ}}}{H_{\frac{N}{2}-1}-H_{l-1}}-1\right)}\right)^2,
\end{equation}
\begin{equation}
\langle\lambda'_8\rangle=\prod _{l=1}^{\frac{N}{2}-2} \left(\frac{P^8_{\text{succ}}e^{a_1  t_1}}{H_{l} \left(e^{a_1  t_1}+\frac{P^8_{\text{succ}}}{H_{l}}-1\right)}\right)^2,
\end{equation}
\begin{equation}
\langle\lambda_9\rangle=\prod _{l=1}^{\frac{N}{2}-1} \left(\frac{P^8_{\text{succ}} e^{a_1  t_1}}{H_{l} \left(e^{a_1  t_1}+\frac{P^8_{\text{succ}}}{H_{l}}-1\right)}\right)^2,
\end{equation}
\begin{equation}
\langle\lambda_{10}\rangle=\left(\frac{P^8_{\text{succ}} e^{a_1  t_1}}{H_{\frac{N}{2}-1} \left(e^{a_1  t_1}+\frac{P^8_{\text{succ}}}{H_{\frac{N}{2}-1}}-1\right)}\right)^2,
\end{equation}
\begin{equation}
\langle\lambda_{11}\rangle=\prod _{l=1}^{\frac{N}{2}-1} \left(\frac{P^1_{\text{succ}} e^{a_1  t_1}}{H_{l} \left(e^{a_1  t_1}+\frac{P^1_{\text{succ}}}{H_{l}}-1\right)}\right)^2,
\end{equation}
\begin{equation}
\langle\lambda_{12}\rangle=\prod _{l=1}^{\frac{N}{2}} \left(\frac{P^1_{\text{succ}} e^{a_0+a_1  t_1}}{\left(H_{\frac{N}{2}}-H_{l-1}\right) \left(e^{a_0+a_1  t_1}+\frac{P^1_{\text{succ}}}{H_{\frac{N}{2}}-H_{l-1}}-1\right)}\right)^2,
\end{equation}
where $H_m$ is the harmonic number ($H_{m}=\sum_{n=1}^m\frac{1}{n}$).

\begin{widetext}
\begin{center}
\begin{table}
\begin{tabular}{ |c|c|c|c| }
 \hline
 & Decoherence Process  & $P^i_{\text{succ}}$ & $t_{\text{step}}$  \\
\hline
1 & Excitation of two electronic spins ($\ket{0}$-levels) &
                                                            &  \\
& and Bell
    measurement between the &$\frac{1}{4} (4-\eta ) \eta$  & $ \frac{d}{c}+t_{\text{prep}}$  \\
 &  photonic modes. Decoherence on a nuclear&  &  \\
& spin in the same NV center.&&\\
\hline
2 & Collective dephasing of the EPL pairs& &  \\
& already generated while waiting for the& $\frac{\eta}{2 (4-\eta )} $ &$2 \left(\frac{d}{c}+t_{\text{prep}}\right)+P^1_{\text{succ}} (S_g+t_{\text{CNOT}})$  \\
&  generation of the remaining pairs &  & \\
&from 1 to $N/2-1$ &  & \\
\hline
3 & Excitation of $\frac{N}{2}$ electronic spins ($\ket{0}$-levels)
    &$\frac{(4-\eta ) \eta ^{\frac{N}{2}-1} \left(\eta ^2-4 \eta +8\right)^{\frac{N}{4}-1}}{2^{N-2}}$ & $\frac{d }{c}+t_{\text{prep}}$ \\
&and $\frac{N}{2}-1$ Bell measurements between the
  & &  \\
&  photonic modes. Decoherence on a nuclear&  & \\
& spin in the same NV center &&\\
\hline
4 & Collective dephasing of the pairs previously& &   \\
& generated during
        the generation of $\frac{N}{2}$ EPL&  & \\
&  pairs. Split into the dephasing
        while one is& $\frac{\eta}{2 (4-\eta )} $  &$2 \left(\frac{d}{c}+t_{\text{prep}}\right)+P^1_{\text{succ}} (S_g+t_{\text{CNOT}})$\\
& accessing the same NV where the qubits &   &\\
&  are stored (4) and while one is accessing the&   &\\
& other NV centers (4') &&\\
\hline
5& Collective dephasing of $\left(\frac{N}{2}-2\right)$ EPL pairs  &
                                                           &\\
&  while they
       are stored waiting for the others &$\frac{\eta}{2 (4-\eta )} $&$2 \left(\frac{d}{c}+t_{\text{prep}}\right)+P^1_{\text{succ}} (S_g+t_{\text{CNOT}})$\\
&to be generated.&&\\
\hline
6& Dephasing of two qubits while $\left(\frac{N}{2}-1\right)$&$\frac{\eta}{2 (4-\eta )} $ & $2 \left(\frac{d}{c}+t_{\text{prep}}\right)+P^1_{\text{succ}} (S_g+t_{\text{CNOT}})$ \\
& EPL pairs are generated   & &\\
 \hline
7 & Collective dephasing of $(N-2)$ qubits & & \\
& while
        $\left(\frac{N}{2}-1\right)$ EPL pairs are generated.&&\\
&  Split
        into the dephasing while one is&$\frac{\eta}{2 (4-\eta )} $ &$2 \left(\frac{d}{c}+t_{\text{prep}}\right)+P^1_{\text{succ}} (S_g+t_{\text{CNOT}})$ \\
& accessing the same NV where the qubits
        & & \\
& are stored (7) and while one is accessing & & \\
&  the other NV centers (7')  & & \\
 \hline
8 & Collective dephasing of $(N-2)$ qubits & & \\
&while
        $\left(\frac{N}{2}-1\right)$ BK pairs are generated. &&\\
&  Split
        into the dephasing while one is& $\frac{\eta^2}{2}$& $ \frac{d}{c}+t_{\text{prep}}$\\
& accessing the same NV center where
        the qubits& & \\
&  are stored (8) and while one is accessing & & \\
& the other NV centers (8')  & & \\
 \hline
9 & Collective dephasing of $\left(\frac{N}{2}-1\right)$ BK  pairs
       & $\frac{\eta^2}{2}$ & $ \frac{d}{c}+t_{\text{prep}}$\\
& while they are generated. & & \\
 \hline
10 & Dephasing of 2 qubits while $\left(\frac{N}{2}-1\right)$ BK
        & $\frac{\eta^2}{2}$ & $ \frac{d}{c}+t_{\text{prep}}$\\
& pairs are generated. & & \\
 \hline
11 & Dephasing of $\left(\frac{N}{2}-1\right)$ non-maximally
         & $\frac{1}{4} (4-\eta ) \eta$&$ \frac{d}{c}+t_{\text{prep}}$ \\
&entangled pairs while they are generated.   & & \\
 \hline
12 & Dephasing of $\frac{N}{2}$ pairs while $\left(\frac{N}{2}-1\right)$ & & \\
& non-maximally entangled pairs are generated. &$\frac{1}{4} (4-\eta ) \eta$ &$ \frac{d}{c}+t_{\text{prep}}$ \\
 \hline
\end{tabular}
\caption{Decoherence processes. In the table we report the six
  processes during which nuclear dephasing occurs, as well as the
  success probability of each process as a function of the
  transmittivity $\eta$ of a single channel, and the corresponding $t_i$.\label{fig:Table}}
\end{table}
\end{center}

\end{widetext}
%\captionof{Table 1.}{ Decoherence processes. In the table we report the six
%  processes during which nuclear dephasing occurs, as well as the
%  success probability of each process as a function of the
%  transmittivity $\eta$ of a single channel, and the corresponding $t_i$.\label{fig:Table}}
The expression of the fidelity  is given by
$F=\frac{1+\langle\lambda\rangle_{\text{tot}}}{2}$, with
$\langle\lambda\rangle_{\text{tot}}=\prod_i \langle\lambda_i\rangle$,
the product is performed over all the dephasing processes occuring
during each protocol.
In the case of the linear and circular protocols 1, both fidelities are 1,
i.e. $F^l_1=1$, and $F^c_1=1$.
The fidelities $F_i^l$ for the linear protocols are the following
\begin{equation}
F^l_2=\frac{1}{2} \left(1+\langle\lambda_1\rangle^N \langle\lambda_2\rangle\right).
\end{equation}
\begin{equation}
F^l_3=\frac{1}{2} \left(1+\langle\lambda_1\rangle^{2N}\langle\lambda_2\rangle^2\langle \lambda'_4\rangle\langle \lambda_4\rangle\right),
\end{equation}
\begin{equation}
F^l_4=\frac{1}{2} \left(1+ \langle\lambda_1\rangle^{2N-2}\langle\lambda_2\rangle\langle\lambda_5\rangle\langle\lambda_6\rangle\langle\lambda_7\rangle\langle\lambda'_7\rangle\right),
\end{equation}
\begin{equation}
F^l_5=\frac{1}{2} \left(1+\langle\lambda_8\rangle \langle\lambda'_8\rangle \langle\lambda_9\rangle \langle\lambda_{10}\rangle\right),
\end{equation}
\begin{equation}
F^l_6=\frac{1}{2} \left(1+ \langle\lambda_1\rangle^{N-2} \langle\lambda_5\rangle \langle\lambda_6\rangle \langle\lambda_7\rangle \langle\lambda'_7\rangle\right),
\end{equation}
The fidelity $F^c_2$ for the circular protocol 2 is
\begin{equation}
F^c_2=\frac{1}{2} \left(1+\langle\lambda_{11}\rangle^2 \langle\lambda_{12}\rangle\right).
\end{equation}

\section{Calculation of the generation times\label{GenRate}}
In this section, we calculate the GHZ generation times per each
protocol. As a first step, we calculate the time required for
successful BK and EPL generation. Let's focus on the BK process. The
total probability of success is $P^8_{\text{succ}}$. The BK generation
time $t_{\text{BK}}$ is given, then, by the sum of all the times
required to perform the procedure divided by $P^8_{\text{succ}}$, i.e.
\begin{equation}
t_{\text{BK}}=\frac{\frac{d}{2 c}+2t_{\text{prep}}+t_X}{P^5_{\text{succ}}},
\end{equation}
where $d$ is the total distance between two nodes, $t_{\text{prep}}$
is the time required to initialize and excite an electronic spin, and
$t_X$ is the time necessary to implement an X rotation.
Concerning the EPL method, two probabilities are involved. One is the
probability of generating a non-maximally entangled pair,
i.e. $P^1_{\text{succ}}$. The other is the probability of success
performing the distillation procedure, i.e. $\frac{2}{(4-\eta)^2}$. The
EPL generation time $t_{\text{EPL}}$ is
\begin{equation}
t_{\text{EPL}}=\left(\frac{2 \left(\frac{d}{c}+t_{\text{prep}}\right)}{P^1_{\text{succ}}}+S_g+t_{\text{CNOT}}\right)\frac{(4-\eta)^2}{2},
\end{equation}
where $S_g$ is the time necessary for swapping the electronic spin and
a nuclear spin, and $t_{\text{CNOT}}$ is the time required for a CNOT
gate. All the generation times are derived in a similar manner. Linear
protocol 1 consists in applying at once $(N-1)$ times the BK procedure, i.e.
\begin{equation}
t^l_1=\frac{\frac{d}{ c}+4 t_{\text{prep}}}{P_{\text{succ}}^{8\,\,\,\,\,N-1}}.
\end{equation}
Linear protocol 2 is equivalent to protocol 1, but $\frac{N}{2}$ BK
procedures are substituted by the generation of EPR pairs. The generation time is
\begin{equation}
t^l_2=\frac{ H_{\frac{N}{2}} t_{\text{EPL}}+\frac{2d}{c}+2 t_{\text{prep}}+t_X}{P_{\text{succ}}^{8\,\,\,\,\,\,\,\frac{N}{2}-1}},
\end{equation}
where $H_{\frac{N}{2}}$ is the harmonic number ($H_{N}=\sum_{n=1}^N\frac{1}{n}$).

Concerning linear protocol 3, the two probabilities involved are
$P^3_{\text{succ}}$ and the probability that the final distillation
procedure succeeds, i.e. $\frac{2^{N-2}}{(4-\eta )^2 \left(\eta ^2-4
    \eta +8\right)^{\frac{N}{2}-2}}$. Thus, the generation time
$t^l_2$ is
\begin{equation}
t^l_3=\frac{(4-\eta )^2 \left(\eta ^2-4 \eta +8\right)^{\frac{N}{2}-2}}{2^{N-2}}
  \left(\frac{H_{\frac{N}{2}} t_{\text{EPL}}+\frac{d}{ c}}{2 P^3_{\text{succ}}}+S_g+t_{\text{CNOT}}\right).
\end{equation}
The linear protocol 4 consists in the generation of $(N-1)$ EPL pairs
in two different steps, followed by an error correction procedure. Hence, we have $t^l_4$ be equal to
\begin{equation}
t^l_4= \left(\left(H_{\frac{N}{2}}+H_{\frac{N}{2}-1}\right) t_{\text{EPL}}+t_{\text{CNOT}}\right).
\end{equation}
Linear protocols 5 and 6 are similar to linear protocol 4, but the
maximal entangled pairs are generated through the BK procedure or a
mix between BK and EPL, respectively. The two generation rates are
\begin{equation}
t^l_5= \left(\left(H_{\frac{N}{2}}+H_{\frac{N}{2}-1}\right) t_{\text{BK}}+t_{\text{CNOT}}+2 t_X\right),
\end{equation}
and
\begin{equation}
t^l_6= \left(H_{\frac{N}{2}} t_{\text{BK}} + H_{\frac{N}{2}-1} t_{\text{EPL}}+t_{\text{CNOT}}+t_X\right).
\end{equation}

Concerning the circular protocols 1 and 2, they have a similar
structure of the corresponding linear protocols and can be derived
similarly. Indeed
\begin{equation}
t^c_1=\frac{2^{N-1}}{\eta^N}\frac{d}{ c}+2 t_{\text{prep}},
\end{equation}
and
\begin{equation}
t^c_2=\frac{1}{2} (4-\eta )^{N} \left(\frac{2 H_{\frac{N}{2}} \left(\frac{d}{ c}+t_{\text{prep}}\right)}{P^1_{\text{succ}}}+S_g+t_{\text{CNOT}}\right).
\end{equation}

\end{document}